\documentclass[reprint,
 amsmath,amssymb,
pra,
]{revtex4-2}
\usepackage{url,hyperref,lineno,microtype,subcaption,listings,xcolor,graphicx,verbatim}
\usepackage{booktabs}

\definecolor{codegreen}{rgb}{0,0.6,0}
\definecolor{codegray}{rgb}{0.5,0.5,0.5}
\definecolor{codepurple}{rgb}{0.58,0,0.82}
\definecolor{backcolour}{rgb}{0.95,0.95,0.95}

\lstdefinestyle{mystyle}{
    backgroundcolor=\color{backcolour},   
    commentstyle=\color{codegreen},
    keywordstyle=\color{magenta},
    numberstyle=\tiny\color{codegray},
    stringstyle=\color{codepurple},
    basicstyle=\ttfamily,
    breakatwhitespace=false,         
    breaklines=true,                 
    captionpos=b,                    
    keepspaces=true,                 
    numbers=left,                    
    numbersep=5pt,                  
    showspaces=false,                
    showstringspaces=false,
    showtabs=false,                  
    tabsize=2
}

\begin{document}
\title[Connecting Scientometrics]{Connecting Scientometrics: Dimensions as a route to broadening context for analyses} 

\author{Simon J Porter}
 \email{s.porter@digital-science.com}
 \affiliation{%
 Digital Science, 6 Briset Street, London, EC1M 5NR, UK
}%
\author{Daniel W Hook}%
\email{daniel@digital-science.com}
\affiliation{%
 Digital Science, 6 Briset Street, London, EC1M 5NR, UK
}%
\affiliation{Centre for Complexity Research, Imperial College London, London, SW7 2AZ, UK}
\affiliation{Department of Physics, Washington University in St Louis, St Louis, Missouri, US.}

\begin{abstract}
Modern cloud-based data infrastructures open new vistas for the deployment of scientometric data into the hands of practitioners.  These infrastructures lower barriers to entry by making data more available and compute capacity more affordable.  In addition, if data are prepared appropriately, with unique identifiers, it is possible to connect many different types of data.  Bringing broader world data into the hands of practitioners (policymakers, strategists and others) who use scientometrics as a tool can extend their capabilities.  These ideas are explored through connecting Dimensions and World Bank data on Google BigQuery to study international collaboration between countries of different economic classification.
\end{abstract}

\maketitle

\section{Introduction}\label{intro}
Until relatively recently, both for academics and practitioners in bibliometrics and scientometrics, it has been challenging to accumulate data with sufficient scale, coverage and accuracy to perform reliable analysis.  In addition, for those who have been able to bring together data with these characteristics, a further challenge has been access to the computational resources needed to perform complex calculations with these data.  

Significant progress has been made in the last ten years both in terms of the number of sources of bibliometric data that are available and to the level of accuracy and granularity of those data.  Sources such as Microsoft Academic Search, Dimensions, and Google Scholar have joined PubMed, Web of Science and Scopus as significant sources for bibliometric analysis, while source such as Unpaywall have extended these datasets with additional, practical data that is highly relevant for research support systems, policy analysis and strategic decision making.

Since its launch in 2018, Dimensions has differentiated itself from other tools in the space on several fronts, two of which are relevant in the context of this article, these are: i) the use of open unique identifiers as a basis for the data held in the data system (and all the consequences of that choice such as the general inclusion of items in the database without additional editorial intervention); and 2) the broadening of the ``fundamental dataset'' available for scientometric analysis by not only including publication and citation data, but also data on grants, clinical trials, patents and policy documents in a single interlinked graph \citep{hook_dimensions_2018, herzog_dimensions_2020}.  The data structures that naturally emerge when taking this approach to creating Dimensions are well suited to the analysis that we will showcase here.

More recently, the data contained in Dimensions has been made available on the Google BigQuery cloud infrastructure. We believe that infrastructures such as these represent a significant opportunity for the adoption and deployment of scientometric analysis. The use of cloud infrastructure achieves two aims: Firstly, it negates the need to implement large, expensive local processing capabilities, providing on-demand access to computational power at a fraction of the cost of local implementations; and, secondly, it facilitates new modes of data sharing, allowing the mixing of public and private datasets in a secure environment.  Both of these facets of cloud compute serve not only to democratise access to analysis \citep{hook_scaling_2021}, but also facilitate an increasingly iterative and real-time relationship with analysis \citep{hook_real-time_2021}.

Research is an intrinsically and increasingly collaborative activity with an increasingly global set of norms and infrastructures \citep{tijssen_collaborations_2011, waltman_globalisation_2011}. Adams demonstrates that research is becoming a more international endeavour  \cite{adams_fourth_2013}. While the centre of this analysis is concerned with the UK, it is clearly more generally applicable and justifies Adams' claim that we are entering a fourth age of research where the normal mode of research is international rather than individual (first age), departmental (second age) or national (third age). 

\begin{figure}
    \centering
    \includegraphics[width=.95\linewidth]{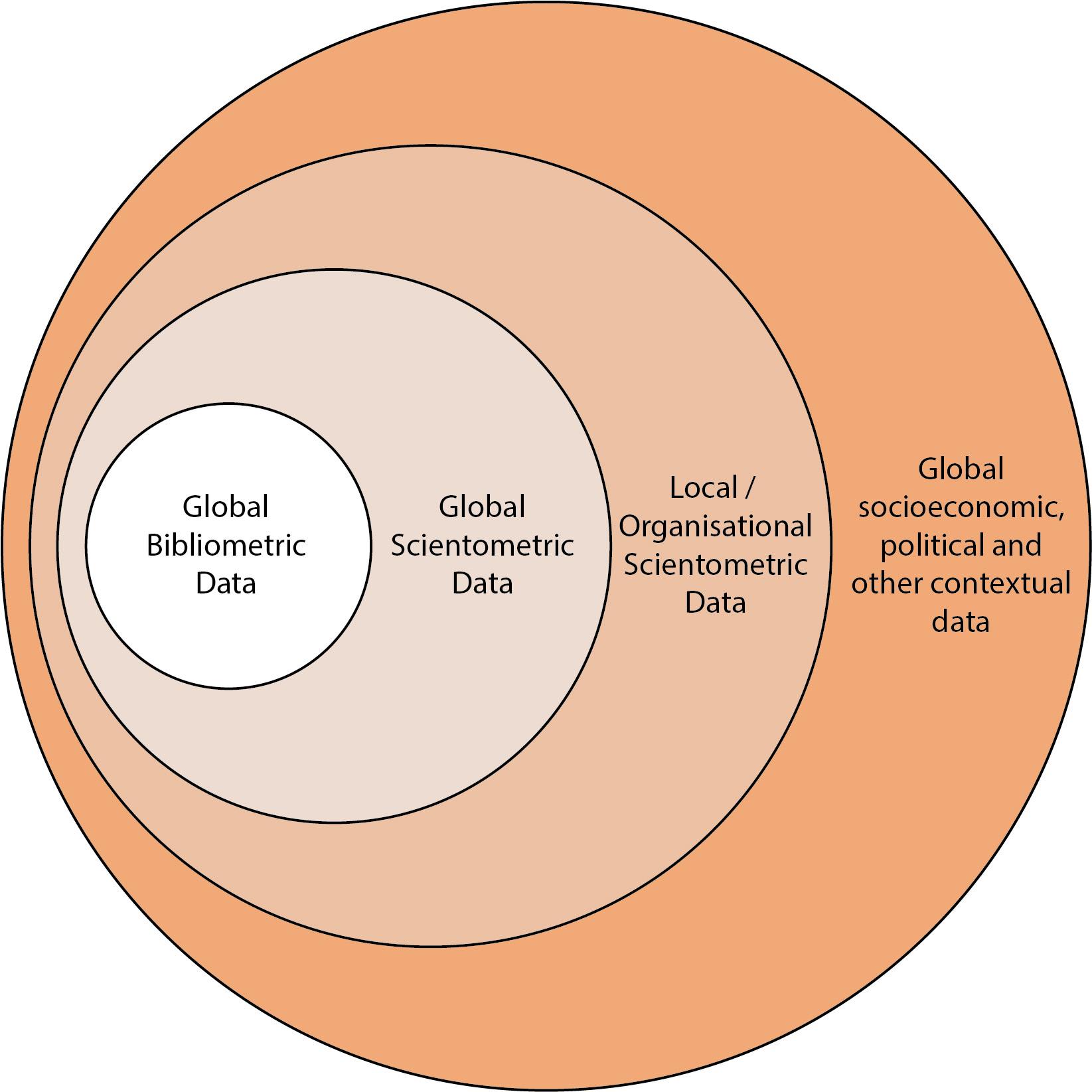}
    \caption{Four modes of data use in scientometrics.}
    \label{fig:fig1}
\end{figure}

In analogy with Adams' ``Fourth age of Research'', we argue that scientometrics has four different modes of data use (see Figure~\ref{fig:fig1}).  The simplest mode is use of global bibliometric data to do high-level analysis such as the construction of benchmarks or the assessment of national, institutional or subject-based research volumes.  The next simplest mode is use of broader data about the research ecosystem - funding data, patent data, and so on in similar use cases to the first mode - benchmarking or general scholarly contextualisation.  The third mode is the use of organisational or local data in analyses: While the first two modes concern global datasets, this mode adds a local reference dataset that is either being analysed in isolation or which is being contextualised using the scholarly datasets and approaches from modes 1 and 2. Third-mode data are often felt to be sensitive in nature and include some of the data held in institutional CRIS or RIMS systems and in funder systems.  These data may including funding success rates, ethnicity data or industrial funding.  The fourth mode concerns broader contextualisation and connection beyond a purely academic or scholarly considerations.  Examples of data and modes in this fourth category include altmetric data (public engagement with research) \citep{sugimoto_scholarly_2017}, socioeconomic data to assess the nature of collaborative trends beyond volume or to track the impact or translation of research.

We assert that policy makers and good strategists naturally attempt to access the fourth mode of scientometric analysis.  There are many examples of researchers who take this broader view - often inspired by or originating from economics as a discipline \cite{lane_assessing_2009, lane_lets_2010, lane_measuring_2011}.  However, getting data in a consistent format and at a level of quality that admits such analyses is challenging and is often the focus of significant research projects in and of itself.  

The approach that we demonstrate in this paper does not negate the challenges of sourcing, curating and manicuring data for quality.  However, it does attempt to showcase a new set of technologies and techniques for re-using data in a broad range of applications and connecting datasets together.  We regard unique identifiers as an enabling infrastructure that allows multiple datasets from different origins to be bought together quickly and easily.  

Our use case, connecting World Bank data as a global economic dataset to Dimensions as a scientometric dataset, is not novel in scope as many other authors have carried out similar studies, see for example \citep{gazni_mapping_2012,chetwood_research_2015}.  Yet, the methodology that we introduce is, we believe, novel in the context of scientometrics and bibliometrics, and demonstrates the potential to perform sophisticated analyses with great speed.

Our motivation for choosing the present example is that collaboration is a core from which many different interesting research questions can be asked. Beyond basic quantification of volumes of collaboration or geographic and institutional loci of collaboration, classification of modes of collaboration is intrinsically interesting.  The Dimensions dataset already contains the data to quantify the attention associated with collaboration (both public attention through Altmetric.com data and scholarly attention from citations), the financial support behind collaboration (grant data), the impact of collaboration (patents, clinical trials, policy documents), and the fields of research being explored in the research.  By connecting to World Bank data we can go beyond these purely scientometric considerations and leverage any of the 1442 indicators that the World Bank makes available on the Google BigQuery data marketplace.

This paper is arranged as follows: In Section~\ref{methods} we give an overview of the computational infrastructure that supports this article, the data infrastructure used for the examples shown and we share basic code listings that highlight the brevity needed to perform these analyses.  In Section~\ref{results} we present some initial results gained using the techniques that are the focus of this paper.  In Section~\ref{discussion} we provide a few thoughts on both the nature of the results and the ease of their production.

\section{Methods}\label{methods}
\subsection{Infrastructure}
Use of cloud infrastructure is at the core of this paper.  While we have chosen to use:
\begin{itemize}
    \item Compute / Data Infrastructure: BigQuery on Google Cloud
    \item Bibliometric/Scientometric data: Dimensions
    \item Programming Language: Python
\end{itemize}

All these choices could be exchanged for equivalents.  The compute and data infrastructure provided by Google has analogues from Amazon, Snowflake, Microsoft, Tencent and others.  Bibliometric and Scientometric data can be sourced from an increasing array of providers.  We chose to use the Google Colaboratory as our development environment and used Python by default.  However, similar analyses can be carried out using other languages and even business intelligence tools such as Tableau.

\subsection{Open Data Standards}
At the core of the methodology of the work reported in this paper are open data standards.  The Dimensions dataset is built on open unique identifiers wherever possible.  When Dimensions was first built, there was no publicly available system of unique identifiers associated with organisations that perform and publish research.  As a result, the Digital Science team created GRID\footnote{http://grid.ac}.  Dimensions on BigQuery includes mappings from each of the research object data types to GRID where an institutional affiliation can be resolved.  

At the time of publication the GRID dataset includes more than 120,000 different research organisations with, among other pieces of metadata, geographical information about the principle campus of each organisation.  GRID attempts to be a gateway to many different standards, which might be helpful when performing analyses, for example: NUTS coding information, geographical longitude and latitude location of the principle campus and mapping to ISO 3166-1 alpha-2 country codes.

More recently Digital Science announced that GRID would not be making any further public releases as the Research Organisation Registry (ROR) has built sufficient support in the community that it has become the principle organisational identifier.  Both GRID and ROR maintain a mappings between the two identifier systems to ensure maximal interoperability.  For simplicity of exposition in the examples shown in this paper, GRID is used, however, RORs could be used at the preference of the researcher.

For the study represented in this paper, the World Bank country income classifications that were updated in July 2021 were used.  Each year these classifications are updated. There are four incoming groups: Low, lower-middle, upper-middle and high-income countries. These classifications relate to the annual Gross National Income (GNI) in USD calculated using the Atlas method exchange rates \citep{world_bank_new_2021}.

At the time of performing this analysis, the classification levels are defined in Table~\ref{table1}. 

\begin{table}[h!]
\begin{center}
\begin{tabular}{cc}\toprule
    \textbf{Group} & \textbf{ GNI per capita (USD)} \\\midrule
    Low income &  $<$1,046\\\midrule
    Lower-middle income &  1,046-4,095 \\\midrule
    Upper-middle income &  4,096-12,695 \\\midrule
    High income &  $>$12,695 \\\bottomrule\hline
\end{tabular}
\end{center}
\caption{Definitions of World Bank income level classifications 2021-2022 \citep{world_bank_new_2021}.}
\label{table1}
\end{table}

For the purposes of this paper, even though we look at data slices over a longer period of time (typically a 10-year period) we have only used the income classifications from 2021.  Our methodology could be improved if we were to track the changes in world bank country classification as publication output changes take place, however, these changes would not significantly effect the outcome of the analysis presented here.  Since our focus is on demonstrating the methodology we have decided that it would be more confusing to present this extension to the analysis than it would be to demonstrate the method using the simpler approach.

\subsection{Data structure}
\begin{figure}[h!]
    \centering
    \includegraphics[width=.95\linewidth]{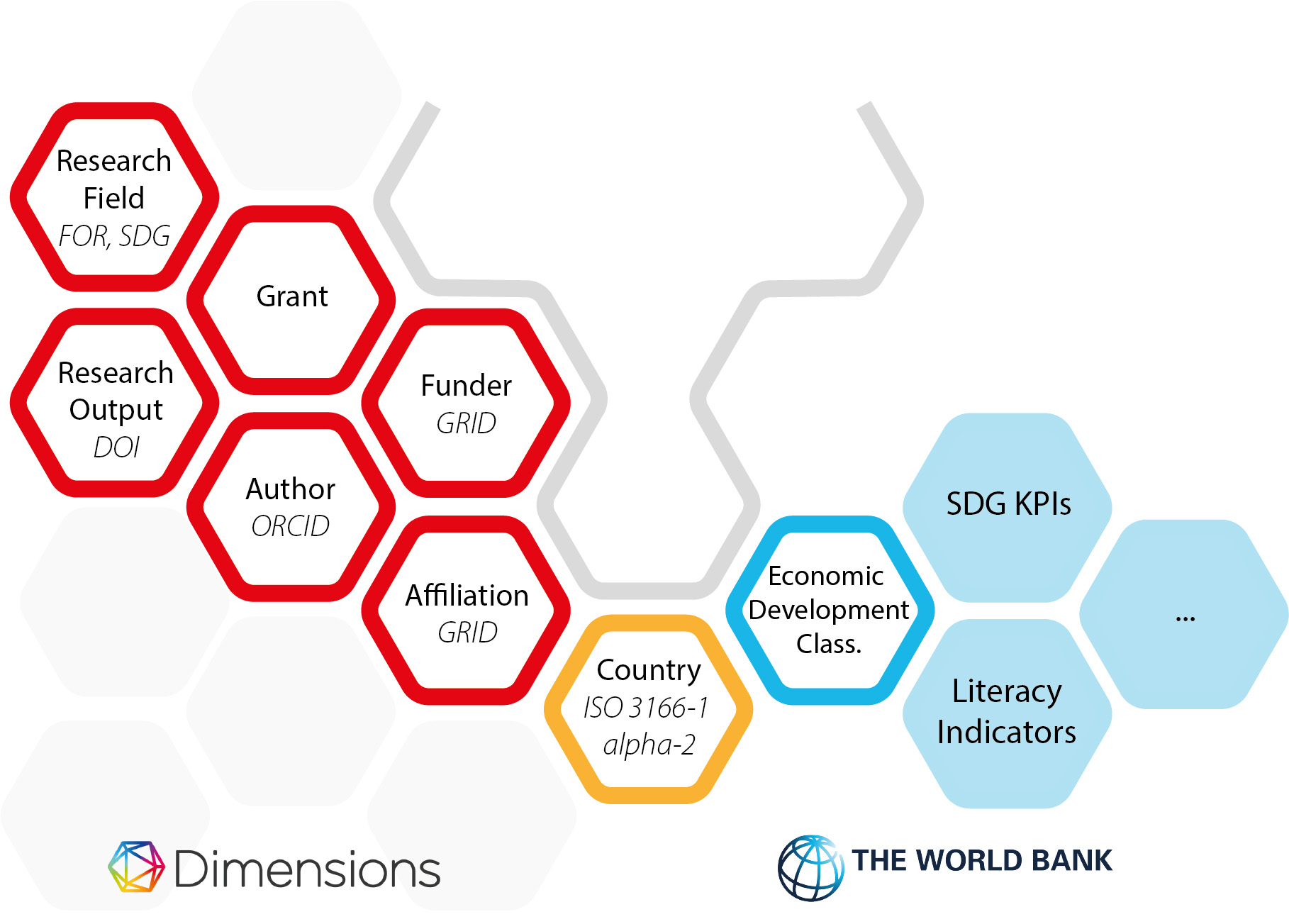}
    \caption{Global collaboration between high income, upper-middle income, lower-middle income and low income countries on publications between 2010 and 2020.}
    \label{fig:fig2}
\end{figure}

Figure~\ref{fig:fig2} shows a schematic representation of two sets of tables in the Google BigQuery environment. Each hexagon is a notional data table in the BigQuery environment apart from the yellow ``country'' hexagon, which has been picked out as the key linking piece of data that connects the Dimensions data (red hexagons) to the World Bank data (blue hexagons).  The solid blue hexagons are examples of other data in the World Bank dataset that are not used in this paper.  The Dimensions data shown here is in the private space of Digital Science (access available via subscription)\footnote{The Dimensions COVID-19 dataset has the same structure as the usual Dimensions data and is freely available on the same infrastructure}.  The World Bank data is publicly available in the BigQuery Marketplace.

In each case, where there is either a unique identifier or public dataset that is a key for one of the Dimensions tables, this has been picked out in italic letters in the hexagon.  For example, researchers in Dimensions are linked to ORCIDs, Research outputs of various types are linked to DOIs from Crossref or Datacite.  Full details of the Dimensions data structure and approach can be found in \cite{hook_dimensions_2018}.

\subsection{Query approach}
The analyses shown in the results section are all based on the same core query, which is shown in Listing~1.  Note that this code is highly economical - the code produces a table listing the number of fully normalised co-authorships between countries by income classification based on publications from the period 2010 to 2020 and also prepares the total number of co-authored international publications in a final column, to allow the calculation of percentages.  The total code (minus comments) is under 30 lines and executes almost immediately on the Google BigQuery infrastructure.

\begin{lstlisting}[language=SQL,caption={Listing to produce an author-contribution-weighted summary of collaborations between countries.},captionpos=b]
\label{lst1}
/* CASE statements select contributions into columns; array_length() is used to normalise contribution for number of co-authors on the paper*/

with  authors as (select p.id, authorder authorder_r, array_length(authors)*array_length(a1.affiliations_address) number_authors
                  from dimensions-ai.data_analytics.publications p,
                         unnest(authors) a1 WITH OFFSET AS authorder
                         
                  where p.year between 2010 and 2020
                  ),

publication_by_proportion as (SELECT distinct  
         p.id,
         authorder,
         affilorder,
         g2.name,
         wb2.income_group,
         aff2.country_code,
         CASE WHEN wb.income_group = 'Low income'          THEN 1/number_authors ELSE null END low_income,
         CASE WHEN wb.income_group = 'Lower middle income' THEN 1/number_authors ELSE null END lower_middle,
         CASE WHEN wb.income_group = 'Upper middle income' THEN 1/number_authors ELSE null END upper_middle_income,
         CASE WHEN wb.income_group = 'High income'         THEN 1/number_authors ELSE null END high_income,
         1/number_authors all_overseas
FROM 
    `dimensions-ai.data_analytics.publications` p,
    /* unnesting by authors and affiliations twice joins the publications table to itself so that we can get the collaborations between authors on each paper and then map to countries*/
        unnest(authors) a1 WITH OFFSET AS authorder,
        unnest(a1.affiliations_address) aff1 WITH OFFSET AS affilorder,
        unnest(authors) a2,
        unnest(a2.affiliations_address) aff2
    inner join authors auth
       on auth.id = p.id
       and authorder = auth.authorder_r
    inner join `dimensions-ai.data_analytics.grid` g1 
       on aff1.grid_id = g1.id
    /* connection to GRID gets us the country data and hence to the ISO codes*/
    inner join `bigquery-public-data.world_bank_wdi.country_summary` wb
        on g1.address.country_code = wb.two_alpha_code 
     inner join `dimensions-ai.data_analytics.grid` g2 
       on aff2.grid_id = g2.id
    inner join `bigquery-public-data.world_bank_wdi.country_summary` wb2
        on g2.address.country_code = wb2.two_alpha_code 
    WHERE 
    /*  restrictions to only count each contribution once; define date range to be 2010 - 2020 */ 
       aff1.country_code != aff2.country_code
       and aff1.grid_id != aff2.grid_id
       and p.year between 2010 and 2020
       and wb2.income_group = 'High income'
)
/*  group results by income group for the aggregation accross all publications*/ 
select name, income_group,country_code, 
       sum(low_income) low_income,
       sum(lower_middle) lower_middle, 
       sum(upper_middle_income) upper_middle_income, 
       sum(high_income) high_income,
       sum(all_overseas) all_overseas
from publication_by_proportion 
group by 1,2,3
order by low_income desc

\end{lstlisting}

\section{Results}\label{results}
In this section, we briefly show some of the results that can be rapidly obtained from the approach that we have described above.  In these examples, we use different facets of the Dimensions data to show some of the capability that is available while only using one of the many tables available in the World Bank dataset.  Thus, these examples merely hint at the opportunity for detailed, real-time analysis.

Each set of results shown below is fruit of taking the coding in Listing~1 and joining on one of the Dimensions tables shown in red in Figure~\ref{fig:fig2}.

\begin{figure}
    \centering
    \includegraphics[width=.95\linewidth]{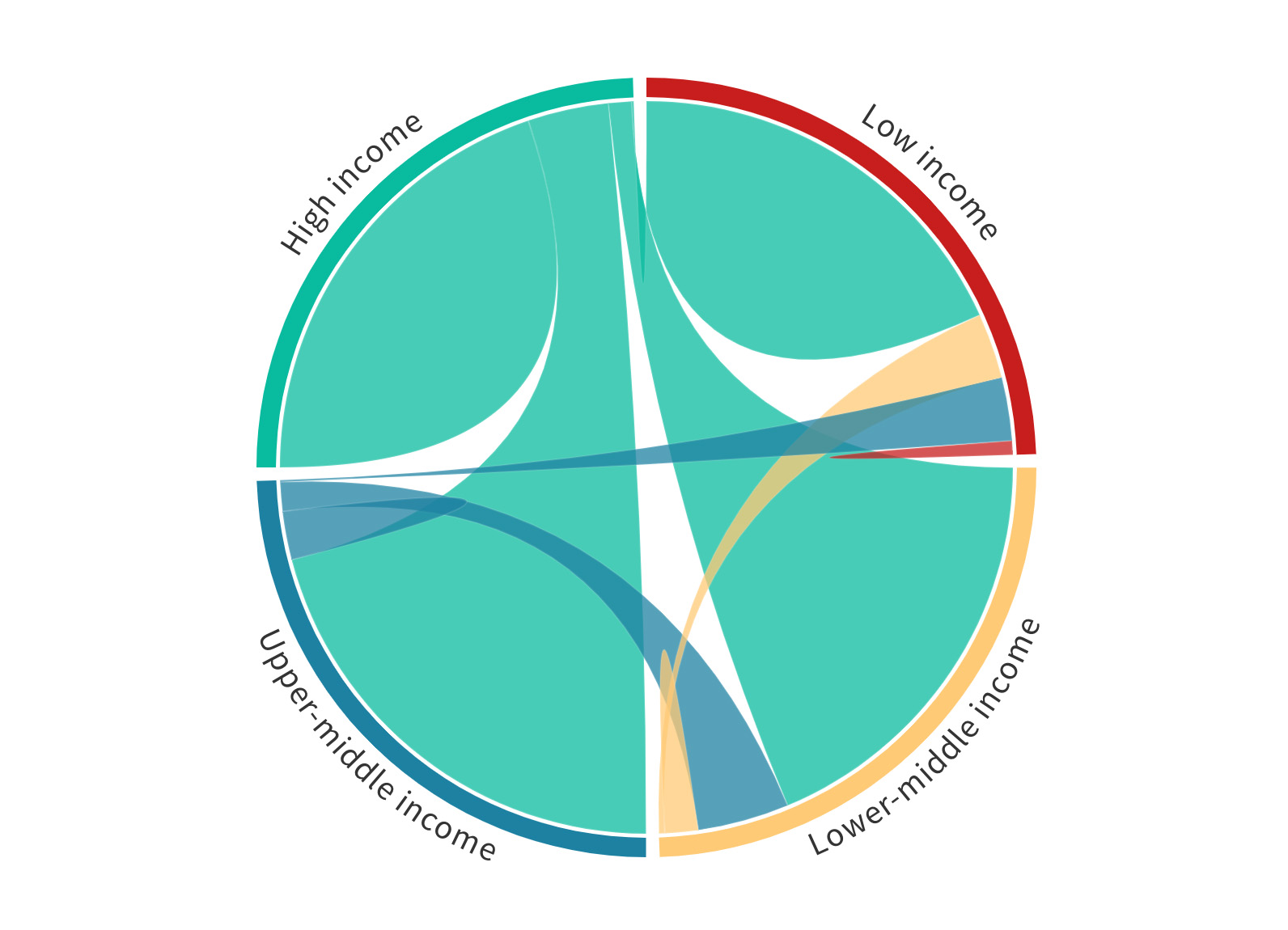}
    \caption{Global collaboration between high income, upper-middle income, lower-middle income and low income countries on publications between 2010 and 2020. Each quadrant in the chord diagram corresponds to 100\% of the research output of countries in each of the brackets.  Thus, Low income levels of output have not been normalised in proportion to lower-middle, upper-middle and high income countries.  In this view we see the proportion collaboration between different economic bands. Values in Table~\ref{table2}.  A chord diagram visualisation is helpful to understand the interplay between bi-lateral relationships but it is important to note that the representation is not entirely faithful as it cannot include multi-lateral relationships.  Hence, if papers are written between participants from three of the four categories or all four categories then they are not represented here.}
    \label{fig:fig3}
\end{figure}

To begin we have calculated the relative levels of collaboration between countries of different World Bank economic classification level based on papers written between 2010 and 2020 (see Figure~\ref{fig:fig3} and Table~\ref{table2}).  In this figure (and the corresponding table) all domestic papers have been removed.  This means that interactions of low-income to low-income countries are interactions between different low-income countries.

\begin{table}[h!]
\begin{center}
\begin{tabular}{lcccc}\toprule
    \textbf{\% of output} & \textbf{Low } & \textbf{Lower-middle } & \textbf{Upper-middle}& \textbf{High } \\\midrule
    Low  &  2.91 & 13.43 &	14.16 &	69.49 \\\midrule
    Lower-middle  &  0.91 &	6.70 &18.46 & 73.92 \\\midrule
    Upper-middle  & 0.38 & 5.31 & 16.58 & 77.70 \\\midrule
    High  &  0.28 & 6.24 & 9.64 & 83.74 \\\bottomrule\hline
\end{tabular}
\end{center}
\caption{Proportion of collaborative output between 2010 and 2020 for global outputs by World Bank country classification level. Rows add to 100\%.}
\label{table2}
\end{table}
Even from this simple analysis it is clear that high-income countries dominate international collaborative relationships with all other economic brackets.  Perhaps more interesting is that there appears to be a somewhat unexpected symmetry in that each of the low, lower-middle, upper-middle, and high income countries all have approximately the proportion between high-income interactions and interactions collectively involving low, lower-middle and upper-middle categories. Table~\ref{table2} shows the detailed proportions shown in Figure~\ref{fig:fig3}.

Perhaps unsurprisingly, the lower the income level of the country, the less engaged they are able to be in the global research community, but it is interesting that low-income countries still make up almost 2.5\% of their collaborative output with other low-income countries.  Another interesting facet shows that upper-middle-income countries are those most engaged with High income countries - presumably because they are beginning to have their own resources as their research economies develop, hence, the engagement is less unidirectional.

\begin{figure}
    \centering
    \includegraphics[width=.95\linewidth]{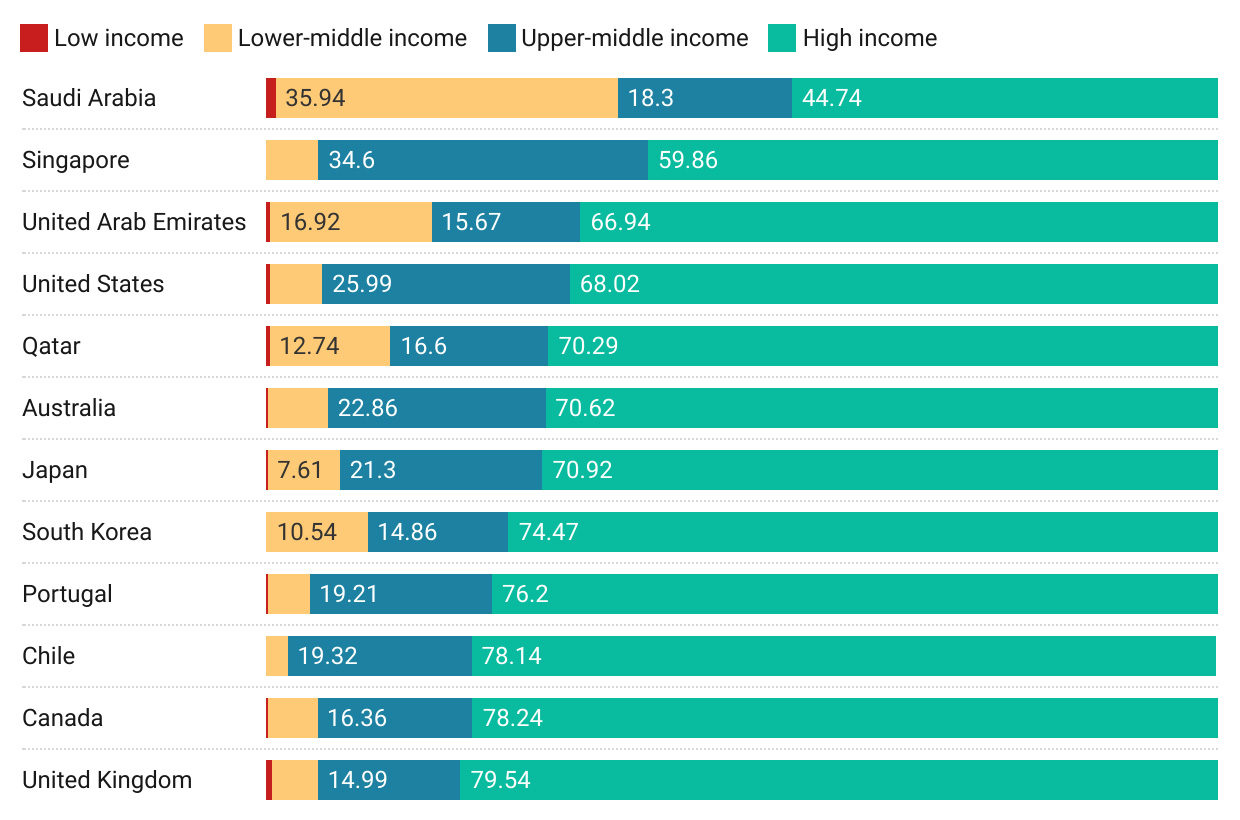}
    \caption{Top 12 high-income countries with the highest non-high-income country collaborations, ordered by amount cumulative proportion of non-high-income collaboration.}
    \label{fig:fig4}
\end{figure}

Having setup this basic framework, many different directions of enquiry are available to us.  For example, we might wish to understand which high-income countries are the most collaborative with countries of lower income levels.  Figure~\ref{fig:fig4} shows the top 12 high-income countries by level of non-high-income country collaboration.  This figure is ordered by the highest level of non-high-income collaboration and hence does not emphasize low-income collaborations.  No specific geographic focus emerges from the high-level data featured in Figure~\ref{fig:fig4}. There may be specific geographic patterns that suggest themselves on further analysis of the data - geographic proximity to countries in lower economic brackets may be one theory.  In the case of the UK and the US, there may be historical or linguistic forces at play - a more detailed study would be needed.  We merely point out that this analysis is a gateway to ask such questions.

\begin{figure}[h!]
    \centering
    \includegraphics[width=.95\linewidth]{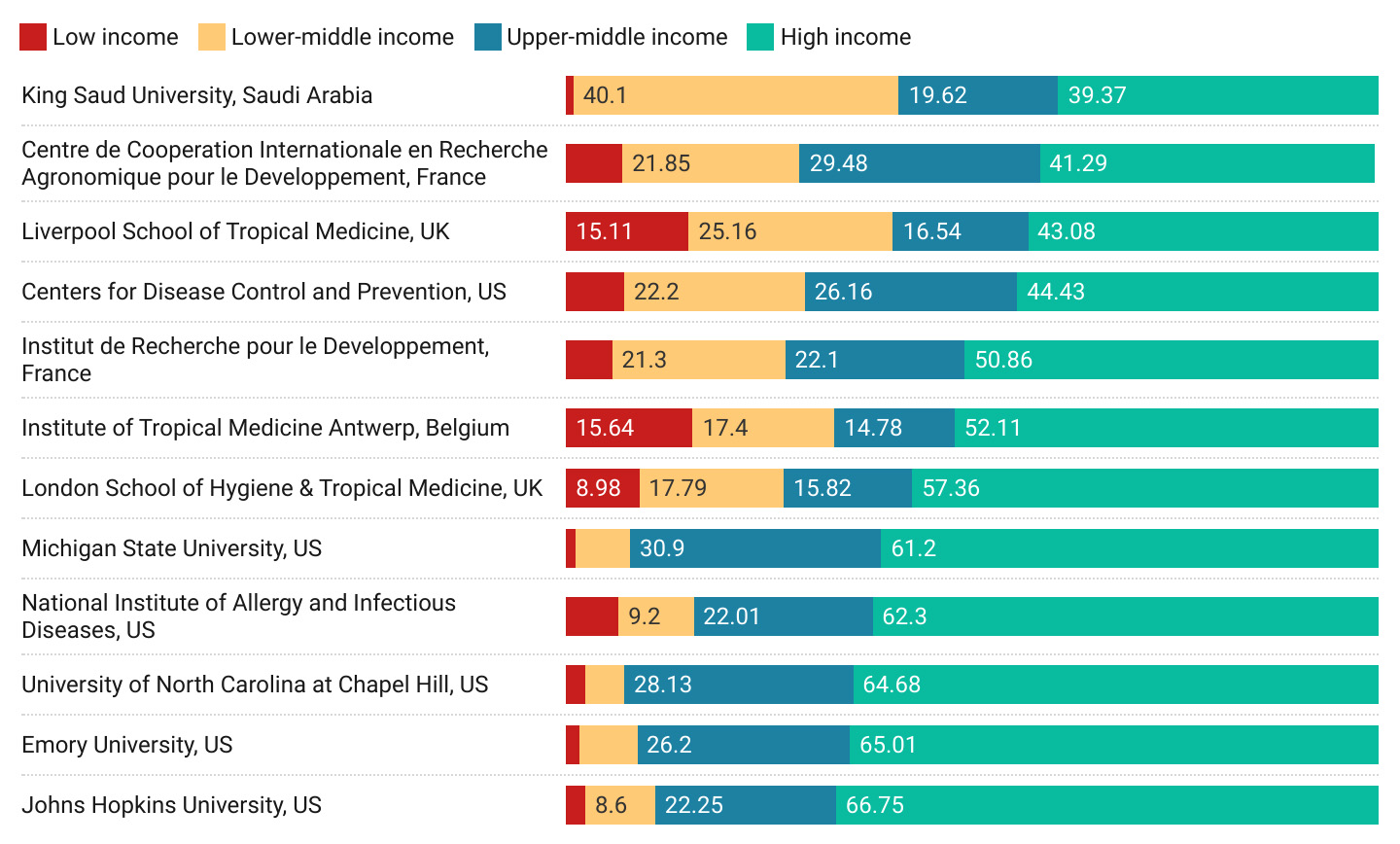}
    \caption{Top 12 High-income-country research institutions with highest non-high-income collaborations.}
    \label{fig:fig5}
\end{figure}

A similar analysis can be performed at an institutional level, which is shown in Figure~\ref{fig:fig5}.  In this figure, almost all the previous countries disappear from the analysis showing that the ability to quickly change between different aggregating objects (countries to institutions) is an important feature of iterative exploratory analysis.  However, this figure suggests an alternative line of enquiry - the names of institutions are suggestive of subject biases that might lead to preferred relationships with developing economies - particularly around medicine and tropical diseases.  

\begin{figure}[h!]
    \centering
    \includegraphics[width=.95\linewidth]{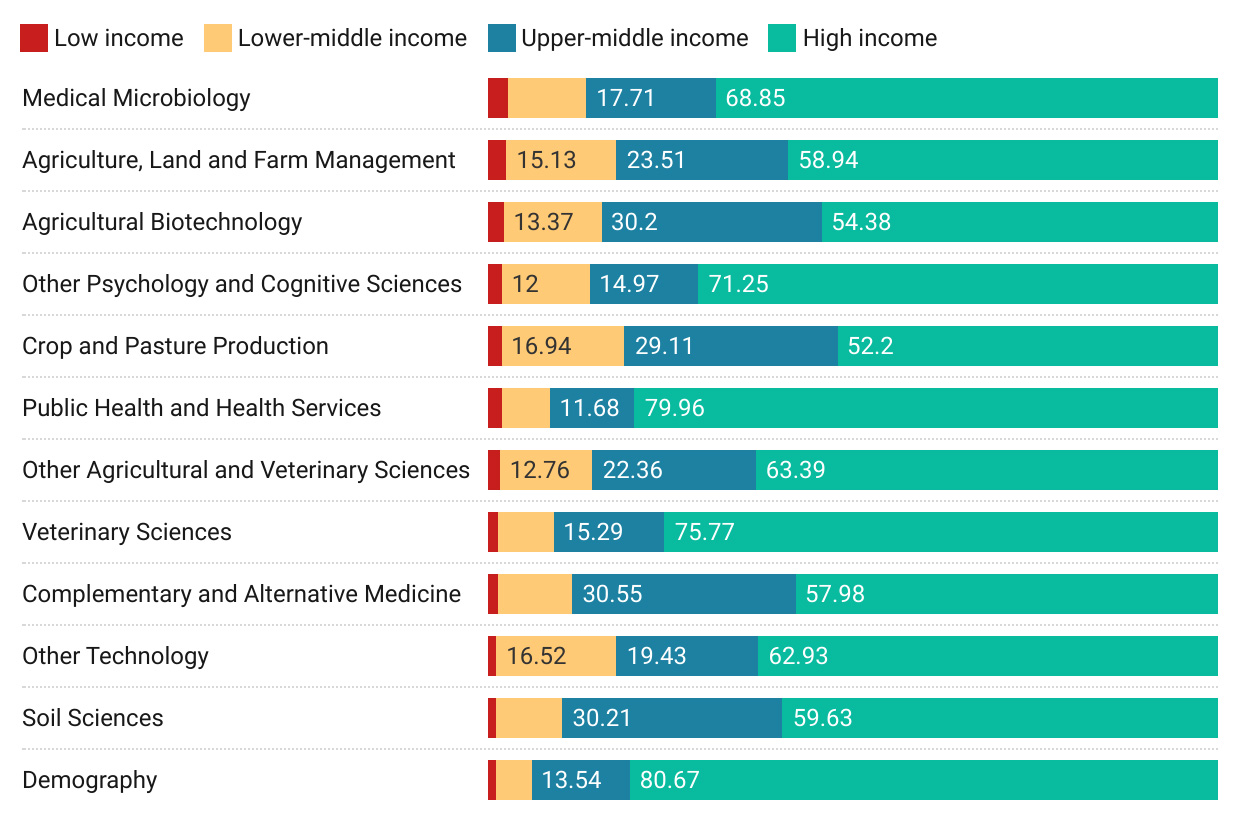}
    \caption{Top 12 high-to-Low-income-country collaborations by field of research ordered by proportion of low-income country collaborations.}
    \label{fig:fig6}
\end{figure}

Figure~\ref{fig:fig6} begins to explore the subject areas that high-income countries collaborate on by economic class using the ANZSRC Field of Research Code classification.  Ordered by participation with Low income countries, this figure suggests that medical,  agricultural and sociological collaborations may be the mainstay of collaborations between high income and low income participants.

\begin{figure}[h!]
    \centering
    \includegraphics[width=.95\linewidth]{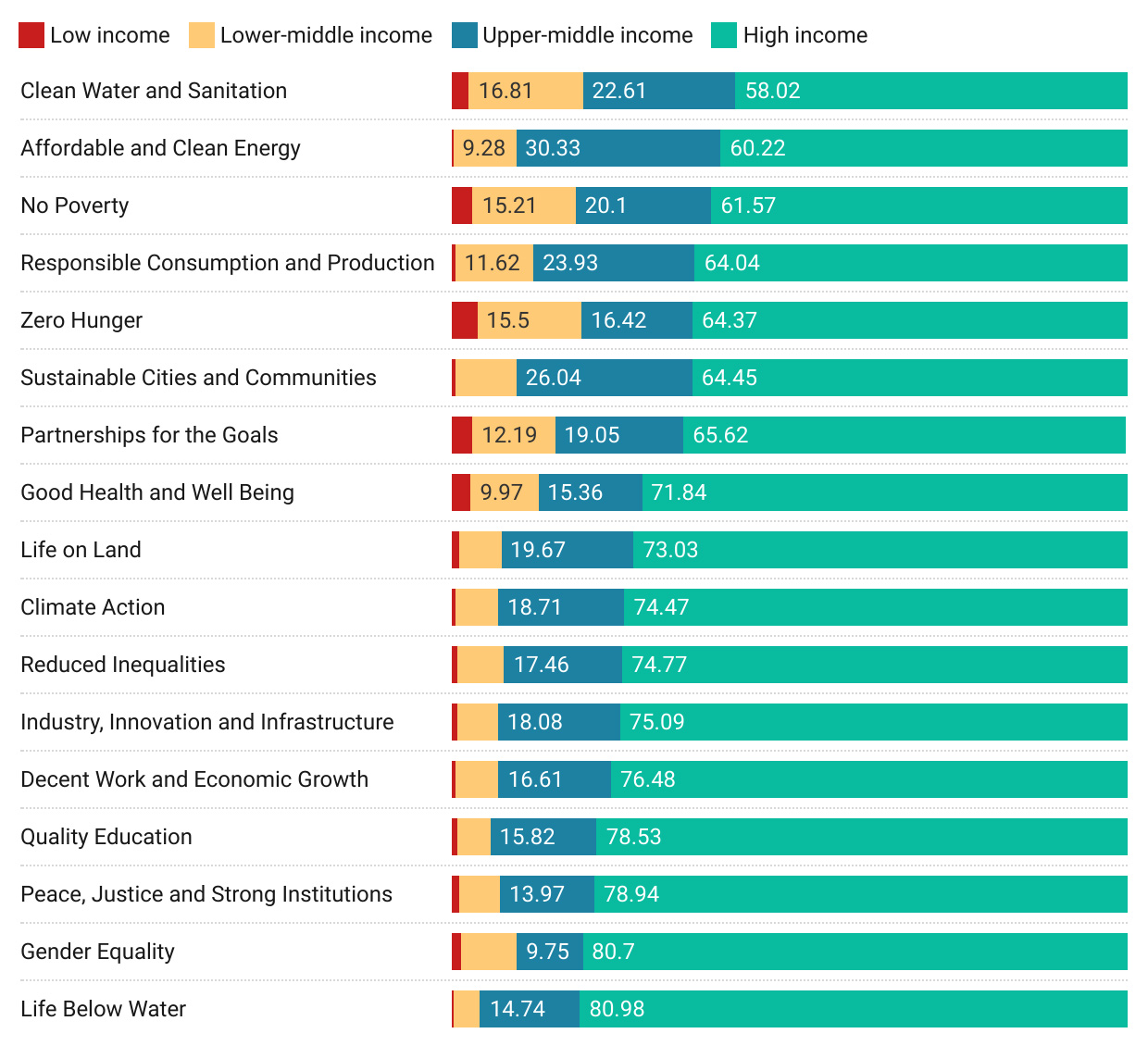}
    \caption{Sustainable development goal collaborations from the perspective of low-income countries, ordered by cumulative non-high-income proportion of research.}
    \label{fig:fig7}
\end{figure}

To add a slightly different perspective, we examine which economic classifications of partner are working with low-income countries on different United Nations Sustainable development goals (SDG).  Figure~\ref{fig:fig7} shows how much collaboration is taking place between low-income countries and other economic levels by SDG.  Hence, we can see that the largest collaboration with higher income countries (as well as with High-income countries specifically) is on the Clean Water and Sanitation SDG.

\begin{figure}[h!]
    \centering
    \includegraphics[width=.95\linewidth]{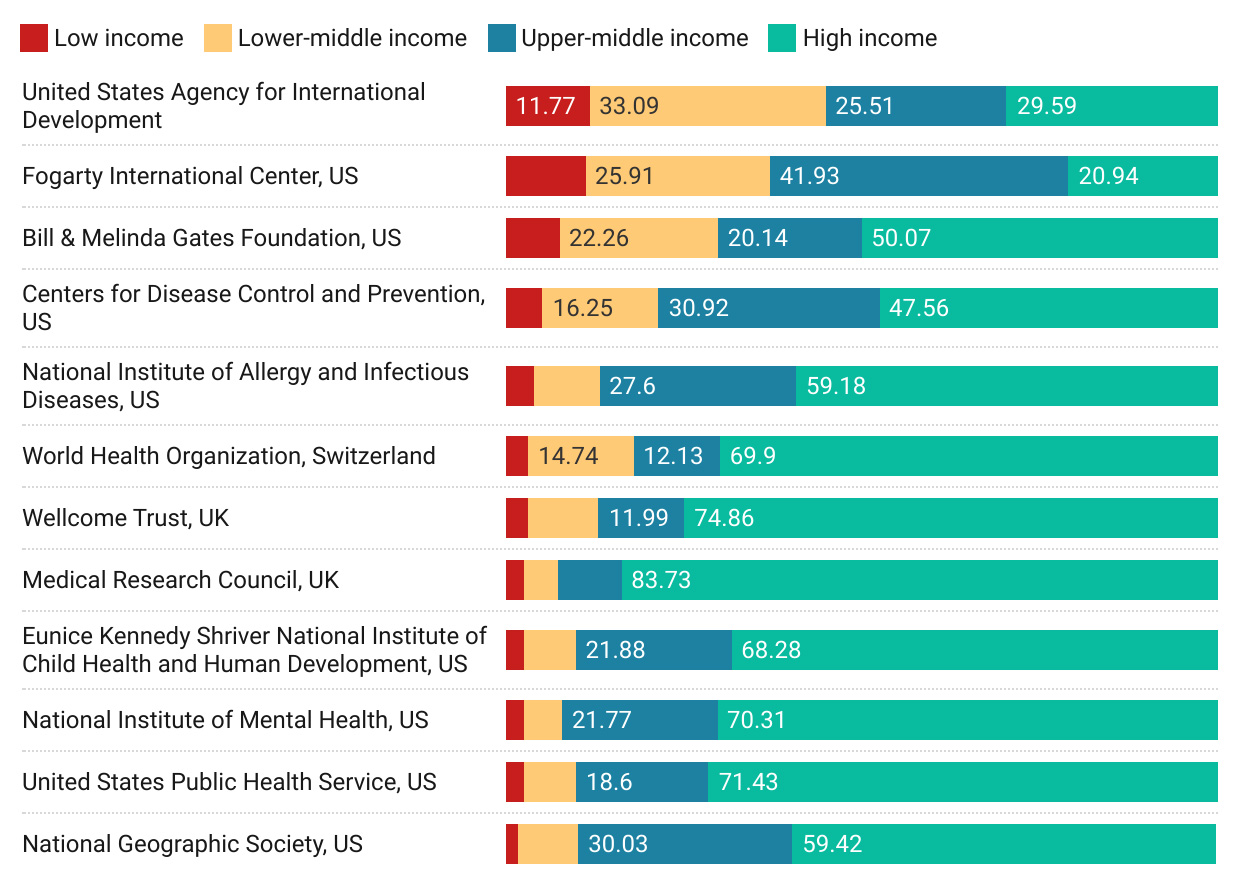}
    \caption{Funding acknowledged in papers co-authored between high-income countries and lower income groups, ordered by proportion of low-income research.}
    \label{fig:fig8}
\end{figure}

Finally, in Figure~\ref{fig:fig8} we report on funding, from the perspective of high-income countries.  The figure shows the worlds funders who have the highest proportion of the international outputs in which they are acknowledged associated with collaborators in countries of lower economic status (ordered by largest proportion of low income collaborations).
\newpage
\section{Discussion}\label{discussion}
\subsection{Discussion of results}
While the main point of this paper is to show the capability of the infrastructure that underlies the results shown here, we start this discussion section with a brief set of observations regarding the analysis in section~\ref{results}.  
In most cultures research is viewed as a positive force that enables economic prosperity and which helps address challenges both practical and intellectual in many different areas from industry to medicine and sociology to art.  At the core of research is collaboration with others.  We have presented an analysis in which we have focused entirely on proportion and not on volume.  This could be considered a weakness of the approach, however, through this lens we can see that, for example,  low income countries choose not to work overly with other low income countries. Even when viewed from the perspective of low income countries (Figure~\ref{fig:fig7}, engagement is relatively lower - and on topics that are important and often highly meaningful to low-income countries.  However, rather than being a question of bias or exclusionary practices the persistence of this results across many lenses suggests systemic issues for low income countries around the fundamental capacity to engage in an international mode: potentially because low income countries have no yet developed sufficient research capacity or formal research infrastructures to engage.

Two figures show a marginally different spectrum of engagement - Figure~\ref{fig:fig5}, for institutions and Figure~\ref{fig:fig8} around funding.  In Figure~\ref{fig:fig5} we see that while it may be difficult for low income countries to engage at a macro level there can be significant collaboration on specific topics with specific institutions.  It is unsurprising that a different collaborative behaviour emerges around disease control, and medicine, as can be seen from the names of the institutions listed in the figure (or the current reputation for medical prowess of the institutions in the list). In Figure~\ref{fig:fig8} we see that funders with specific missions are also successful at directing their funding towards supporting collaboration between high income and low income countries.

\subsection{Reflections on infrastructure}
In this short paper, we have attempted to show the potential for the use of modern technical infrastructures in bibliometric and scientometric analysis and we have suggested a framework to think about the different modes of data usage.  Much that we have demonstrated may be of interest to the research community as a means to translate the technologies and approaches that they develop to be used broadly by practitioners. The enabling infrastructures (such as unique identifiers) that are described here are well known in both the research context and the practitioner context, yet the allied enabling platforms such as Google BigQuery are, as yet, less well used.  The exciting opportunity for researchers is the datasets that they create as part of their research projects may gain additional use and generate additional value to the broader academic community if they are constructed with good use of open data standard and can be shared on infrastructures such as those described here.  

The beneficiaries of the technologies described here are manifold. Research policy makers, strategists and decision makers need to be able to bring their analyses in contact with broader global considerations such as those from financial, economic and sociological sectors.  The technology approach makes that more tractable for those who lack significant resources (to build their own compute and data handling infrastructure).

The notion that research should exist in a broader context and that in order to do that it needs access to well structured ``computable data'' is not a new one \citep{wolfram_making_2010}.  The Mathematica and Wolfram Language system has created an ecosystem where researchers can call upon well-structured contextualising statistical content has been in place for some time, having been introduced in Mathematica in 2007 \citep{wolfram_research_countrydatawolfram_2021}.  The team has gone on to include domain-specific data in a dizzying array of different fields.  However, this ecosystem does not allow the easy addition of new data sources and requires users to be comfortable with the Wolfram Language.  While the data in Google BigQuery and other similar environments may be less well structured than the data in the Wolfram environment, it is more easy to contribute to leverage a mix of private and public datasets through open identifier schemes and, due to the nature of the underlying technology approach, these new platforms are all but language agnostic.

Finally, we reflect that we have not seen extensive analysis that focus on the issues highlighted by the analysis that we performed as an example in this paper in the general bibliometric or scientometric literature and believe that a significant detailed study is required that moves beyond collaboration volume or attention to the modes of collaboration highlighted in our brief analysis here.  We believe that there is a rich seam of data that can no easily be explored through the techniques that we have shared in this paper.  

While we have focused in this paper on one single facet of World Bank data, there are many other tables within the World Bank dataset that can be explored.  There are also many other datasets that have been prepared to be used in the ways described in this article.  Google's cloud marketplace already includes public datasets from the Centers for Disease Control, the Broad Institute, the United States Census Bureau and many others.

Analyses such as these have the capacity to empower practitioners to highlight and address issues of significant importance beyond purely scholarly issues.  Connecting scientometrics to broader datasets using the types of methods shown here gives policy makers at all levels a set of invaluable, accessible tools to make better cases for support, and to make better decisions.

\section*{Conflict of Interest Statement}
All co-authors of this paper are employees of Digital Science, the creator and provider of \emph{Dimensions}.

\section*{Author Contributions}
Ideas for this article were generated and refined by SJP and DWH. Data analysis for this article was performed by SJP. Interpretation of analysis was performed by all co-authors. The article was drafted by DWH and all authors collaborated on editing the article and responding to referee comments.


\section*{Data Availability Statement}
The datasets analysed for this study can be found in the Figshare repository at \url{http://10.6084/m9.figshare.17197262}.

\bibliography{frontiers.bib}

\begin{thebibliography}{18}%
\makeatletter
\providecommand \@ifxundefined [1]{%
 \@ifx{#1\undefined}
}%
\providecommand \@ifnum [1]{%
 \ifnum #1\expandafter \@firstoftwo
 \else \expandafter \@secondoftwo
 \fi
}%
\providecommand \@ifx [1]{%
 \ifx #1\expandafter \@firstoftwo
 \else \expandafter \@secondoftwo
 \fi
}%
\providecommand \natexlab [1]{#1}%
\providecommand \enquote  [1]{``#1''}%
\providecommand \bibnamefont  [1]{#1}%
\providecommand \bibfnamefont [1]{#1}%
\providecommand \citenamefont [1]{#1}%
\providecommand \href@noop [0]{\@secondoftwo}%
\providecommand \href [0]{\begingroup \@sanitize@url \@href}%
\providecommand \@href[1]{\@@startlink{#1}\@@href}%
\providecommand \@@href[1]{\endgroup#1\@@endlink}%
\providecommand \@sanitize@url [0]{\catcode `\\12\catcode `\$12\catcode
  `\&12\catcode `\#12\catcode `\^12\catcode `\_12\catcode `\%12\relax}%
\providecommand \@@startlink[1]{}%
\providecommand \@@endlink[0]{}%
\providecommand \url  [0]{\begingroup\@sanitize@url \@url }%
\providecommand \@url [1]{\endgroup\@href {#1}{\urlprefix }}%
\providecommand \urlprefix  [0]{URL }%
\providecommand \Eprint [0]{\href }%
\providecommand \doibase [0]{https://doi.org/}%
\providecommand \selectlanguage [0]{\@gobble}%
\providecommand \bibinfo  [0]{\@secondoftwo}%
\providecommand \bibfield  [0]{\@secondoftwo}%
\providecommand \translation [1]{[#1]}%
\providecommand \BibitemOpen [0]{}%
\providecommand \bibitemStop [0]{}%
\providecommand \bibitemNoStop [0]{.\EOS\space}%
\providecommand \EOS [0]{\spacefactor3000\relax}%
\providecommand \BibitemShut  [1]{\csname bibitem#1\endcsname}%
\let\auto@bib@innerbib\@empty
\bibitem [{\citenamefont {Hook}\ \emph {et~al.}(2018)\citenamefont {Hook},
  \citenamefont {Porter},\ and\ \citenamefont {Herzog}}]{hook_dimensions_2018}%
  \BibitemOpen
  \bibfield  {author} {\bibinfo {author} {\bibfnamefont {D.~W.}\ \bibnamefont
  {Hook}}, \bibinfo {author} {\bibfnamefont {S.~J.}\ \bibnamefont {Porter}},\
  and\ \bibinfo {author} {\bibfnamefont {C.}~\bibnamefont {Herzog}},\
  }\bibfield  {title} {\bibinfo {title} {Dimensions: {Building} {Context} for
  {Search} and {Evaluation}},\ }\href {https://doi.org/10.3389/frma.2018.00023}
  {\bibfield  {journal} {\bibinfo  {journal} {Frontiers in Research Metrics and
  Analytics}\ }\textbf {\bibinfo {volume} {3}},\ \bibinfo {pages} {23}
  (\bibinfo {year} {2018})}\BibitemShut {NoStop}%
\bibitem [{\citenamefont {Herzog}\ \emph {et~al.}(2020)\citenamefont {Herzog},
  \citenamefont {Hook},\ and\ \citenamefont
  {Konkiel}}]{herzog_dimensions_2020}%
  \BibitemOpen
  \bibfield  {author} {\bibinfo {author} {\bibfnamefont {C.}~\bibnamefont
  {Herzog}}, \bibinfo {author} {\bibfnamefont {D.}~\bibnamefont {Hook}},\ and\
  \bibinfo {author} {\bibfnamefont {S.}~\bibnamefont {Konkiel}},\ }\bibfield
  {title} {\bibinfo {title} {Dimensions: {Bringing} down barriers between
  scientometricians and data},\ }\href {https://doi.org/10.1162/qss_a_00020}
  {\bibfield  {journal} {\bibinfo  {journal} {Quantitative Science Studies}\
  }\textbf {\bibinfo {volume} {1}},\ \bibinfo {pages} {387} (\bibinfo {year}
  {2020})}\BibitemShut {NoStop}%
\bibitem [{\citenamefont {Hook}\ and\ \citenamefont
  {Porter}(2021)}]{hook_scaling_2021}%
  \BibitemOpen
  \bibfield  {author} {\bibinfo {author} {\bibfnamefont {D.~W.}\ \bibnamefont
  {Hook}}\ and\ \bibinfo {author} {\bibfnamefont {S.~J.}\ \bibnamefont
  {Porter}},\ }\bibfield  {title} {\bibinfo {title} {Scaling {Scientometrics}:
  {Dimensions} on {Google} {BigQuery} as an {Infrastructure} for
  {Large}-{Scale} {Analysis}},\ }\href
  {https://doi.org/10.3389/frma.2021.656233} {\bibfield  {journal} {\bibinfo
  {journal} {Frontiers in Research Metrics and Analytics}\ }\textbf {\bibinfo
  {volume} {6}},\ \bibinfo {pages} {5} (\bibinfo {year} {2021})}\BibitemShut
  {NoStop}%
\bibitem [{\citenamefont {Hook}\ \emph {et~al.}(2021)\citenamefont {Hook},
  \citenamefont {Porter}, \citenamefont {Draux},\ and\ \citenamefont
  {Herzog}}]{hook_real-time_2021}%
  \BibitemOpen
  \bibfield  {author} {\bibinfo {author} {\bibfnamefont {D.~W.}\ \bibnamefont
  {Hook}}, \bibinfo {author} {\bibfnamefont {S.~J.}\ \bibnamefont {Porter}},
  \bibinfo {author} {\bibfnamefont {H.}~\bibnamefont {Draux}},\ and\ \bibinfo
  {author} {\bibfnamefont {C.~T.}\ \bibnamefont {Herzog}},\ }\bibfield  {title}
  {\bibinfo {title} {Real-{Time} {Bibliometrics}: {Dimensions} as a {Resource}
  for {Analyzing} {Aspects} of {COVID}-19},\ }\href
  {https://doi.org/10.3389/frma.2020.595299} {\bibfield  {journal} {\bibinfo
  {journal} {Frontiers in Research Metrics and Analytics}\ }\textbf {\bibinfo
  {volume} {5}},\ \bibinfo {pages} {25} (\bibinfo {year} {2021})}\BibitemShut
  {NoStop}%
\bibitem [{\citenamefont {Tijssen}\ \emph {et~al.}(2011)\citenamefont
  {Tijssen}, \citenamefont {Waltman},\ and\ \citenamefont {van
  Eck}}]{tijssen_collaborations_2011}%
  \BibitemOpen
  \bibfield  {author} {\bibinfo {author} {\bibfnamefont {R.~J.~W.}\
  \bibnamefont {Tijssen}}, \bibinfo {author} {\bibfnamefont {L.}~\bibnamefont
  {Waltman}},\ and\ \bibinfo {author} {\bibfnamefont {N.~J.}\ \bibnamefont {van
  Eck}},\ }\bibfield  {title} {\bibinfo {title} {Collaborations span 1,553
  kilometres},\ }\href {https://doi.org/10.1038/473154a} {\bibfield  {journal}
  {\bibinfo  {journal} {Nature}\ }\textbf {\bibinfo {volume} {473}},\ \bibinfo
  {pages} {154} (\bibinfo {year} {2011})}\BibitemShut {NoStop}%
\bibitem [{\citenamefont {Waltman}\ \emph {et~al.}(2011)\citenamefont
  {Waltman}, \citenamefont {Tijssen},\ and\ \citenamefont
  {Eck}}]{waltman_globalisation_2011}%
  \BibitemOpen
  \bibfield  {author} {\bibinfo {author} {\bibfnamefont {L.}~\bibnamefont
  {Waltman}}, \bibinfo {author} {\bibfnamefont {R.~J.~W.}\ \bibnamefont
  {Tijssen}},\ and\ \bibinfo {author} {\bibfnamefont {N.~J.~v.}\ \bibnamefont
  {Eck}},\ }\bibfield  {title} {\bibinfo {title} {Globalisation of science in
  kilometres},\ }\href {https://doi.org/10.1016/j.joi.2011.05.003} {\bibfield
  {journal} {\bibinfo  {journal} {Journal of Informetrics}\ }\textbf {\bibinfo
  {volume} {5}},\ \bibinfo {pages} {574} (\bibinfo {year} {2011})}\BibitemShut
  {NoStop}%
\bibitem [{\citenamefont {Adams}(2013)}]{adams_fourth_2013}%
  \BibitemOpen
  \bibfield  {author} {\bibinfo {author} {\bibfnamefont {J.}~\bibnamefont
  {Adams}},\ }\bibfield  {title} {\bibinfo {title} {The fourth age of
  research},\ }\href {https://doi.org/10.1038/497557a} {\bibfield  {journal}
  {\bibinfo  {journal} {Nature}\ }\textbf {\bibinfo {volume} {497}},\ \bibinfo
  {pages} {557} (\bibinfo {year} {2013})}\BibitemShut {NoStop}%
\bibitem [{\citenamefont {Sugimoto}\ \emph {et~al.}(2017)\citenamefont
  {Sugimoto}, \citenamefont {Work}, \citenamefont {Larivière},\ and\
  \citenamefont {Haustein}}]{sugimoto_scholarly_2017}%
  \BibitemOpen
  \bibfield  {author} {\bibinfo {author} {\bibfnamefont {C.~R.}\ \bibnamefont
  {Sugimoto}}, \bibinfo {author} {\bibfnamefont {S.}~\bibnamefont {Work}},
  \bibinfo {author} {\bibfnamefont {V.}~\bibnamefont {Larivière}},\ and\
  \bibinfo {author} {\bibfnamefont {S.}~\bibnamefont {Haustein}},\ }\bibfield
  {title} {\bibinfo {title} {Scholarly use of social media and altmetrics: {A}
  review of the literature},\ }\href {https://doi.org/10.1002/asi.23833}
  {\bibfield  {journal} {\bibinfo  {journal} {Journal of the Association for
  Information Science and Technology}\ }\textbf {\bibinfo {volume} {68}},\
  \bibinfo {pages} {2037} (\bibinfo {year} {2017})},\ \bibinfo {note}
  {\_eprint:
  https://onlinelibrary.wiley.com/doi/pdf/10.1002/asi.23833}\BibitemShut
  {NoStop}%
\bibitem [{\citenamefont {Lane}(2009)}]{lane_assessing_2009}%
  \BibitemOpen
  \bibfield  {author} {\bibinfo {author} {\bibfnamefont {J.}~\bibnamefont
  {Lane}},\ }\bibfield  {title} {\bibinfo {title} {Assessing the {Impact} of
  {Science} {Funding}},\ }\href {https://doi.org/10.1126/science.1175335}
  {\bibfield  {journal} {\bibinfo  {journal} {Science}\ }\textbf {\bibinfo
  {volume} {324}},\ \bibinfo {pages} {1273} (\bibinfo {year} {2009})},\
  \bibinfo {note} {publisher: American Association for the Advancement of
  Science}\BibitemShut {NoStop}%
\bibitem [{\citenamefont {Lane}(2010)}]{lane_lets_2010}%
  \BibitemOpen
  \bibfield  {author} {\bibinfo {author} {\bibfnamefont {J.}~\bibnamefont
  {Lane}},\ }\bibfield  {title} {\bibinfo {title} {Let's make science metrics
  more scientific},\ }\href {https://doi.org/10.1038/464488a} {\bibfield
  {journal} {\bibinfo  {journal} {Nature}\ }\textbf {\bibinfo {volume} {464}},\
  \bibinfo {pages} {488} (\bibinfo {year} {2010})}\BibitemShut {NoStop}%
\bibitem [{\citenamefont {Lane}\ and\ \citenamefont
  {Bertuzzi}(2011)}]{lane_measuring_2011}%
  \BibitemOpen
  \bibfield  {author} {\bibinfo {author} {\bibfnamefont {J.}~\bibnamefont
  {Lane}}\ and\ \bibinfo {author} {\bibfnamefont {S.}~\bibnamefont
  {Bertuzzi}},\ }\bibfield  {title} {\bibinfo {title} {Measuring the {Results}
  of {Science} {Investments}},\ }\href
  {https://doi.org/10.1126/science.1201865} {\bibfield  {journal} {\bibinfo
  {journal} {Science}\ }\textbf {\bibinfo {volume} {331}},\ \bibinfo {pages}
  {678} (\bibinfo {year} {2011})},\ \bibinfo {note} {publisher: American
  Association for the Advancement of Science}\BibitemShut {NoStop}%
\bibitem [{\citenamefont {Gazni}\ \emph {et~al.}(2012)\citenamefont {Gazni},
  \citenamefont {Sugimoto},\ and\ \citenamefont
  {Didegah}}]{gazni_mapping_2012}%
  \BibitemOpen
  \bibfield  {author} {\bibinfo {author} {\bibfnamefont {A.}~\bibnamefont
  {Gazni}}, \bibinfo {author} {\bibfnamefont {C.~R.}\ \bibnamefont
  {Sugimoto}},\ and\ \bibinfo {author} {\bibfnamefont {F.}~\bibnamefont
  {Didegah}},\ }\bibfield  {title} {\bibinfo {title} {Mapping world scientific
  collaboration: {Authors}, institutions, and countries},\ }\href
  {https://doi.org/10.1002/asi.21688} {\bibfield  {journal} {\bibinfo
  {journal} {Journal of the American Society for Information Science and
  Technology}\ }\textbf {\bibinfo {volume} {63}},\ \bibinfo {pages} {323}
  (\bibinfo {year} {2012})}\BibitemShut {NoStop}%
\bibitem [{\citenamefont {Chetwood}\ \emph {et~al.}(2015)\citenamefont
  {Chetwood}, \citenamefont {Ladep},\ and\ \citenamefont
  {Taylor-Robinson}}]{chetwood_research_2015}%
  \BibitemOpen
  \bibfield  {author} {\bibinfo {author} {\bibfnamefont {J.~D.}\ \bibnamefont
  {Chetwood}}, \bibinfo {author} {\bibfnamefont {N.~G.}\ \bibnamefont
  {Ladep}},\ and\ \bibinfo {author} {\bibfnamefont {S.~D.}\ \bibnamefont
  {Taylor-Robinson}},\ }\bibfield  {title} {\bibinfo {title} {Research
  partnerships between high and low-income countries: are international
  partnerships always a good thing?},\ }\href
  {https://doi.org/10.1186/s12910-015-0030-z} {\bibfield  {journal} {\bibinfo
  {journal} {BMC Medical Ethics}\ }\textbf {\bibinfo {volume} {16}},\ \bibinfo
  {pages} {36} (\bibinfo {year} {2015})}\BibitemShut {NoStop}%
\bibitem [{Note1()}]{Note1}%
  \BibitemOpen
  \bibinfo {note} {Http://grid.ac}\BibitemShut {NoStop}%
\bibitem [{\citenamefont {{World Bank}}(2021)}]{world_bank_new_2021}%
  \BibitemOpen
  \bibfield  {author} {\bibinfo {author} {\bibnamefont {{World Bank}}},\ }\href
  {https://blogs.worldbank.org/opendata/new-world-bank-country-classifications-income-level-2021-2022}
  {\bibinfo {title} {New {World} {Bank} country classifications by income
  level: 2021-2022}} (\bibinfo {year} {2021})\BibitemShut {NoStop}%
\bibitem [{Note2()}]{Note2}%
  \BibitemOpen
  \bibinfo {note} {The Dimensions COVID-19 dataset has the same structure as
  the usual Dimensions data and is freely available on the same
  infrastructure}\BibitemShut {NoStop}%
\bibitem [{\citenamefont {Wolfram}(2010)}]{wolfram_making_2010}%
  \BibitemOpen
  \bibfield  {author} {\bibinfo {author} {\bibfnamefont {S.}~\bibnamefont
  {Wolfram}},\ }\bibfield  {title} {\bibinfo {title} {Making the {World}’s
  {Data} {Computable}}\ }(\bibinfo {year} {2010})\BibitemShut {NoStop}%
\bibitem [{\citenamefont {{Wolfram
  Research}}(2021)}]{wolfram_research_countrydatawolfram_2021}%
  \BibitemOpen
  \bibfield  {author} {\bibinfo {author} {\bibnamefont {{Wolfram Research}}},\
  }\href {https://reference.wolfram.com/language/ref/CountryData.html}
  {\bibinfo {title} {{CountryData}—{Wolfram} {Language} {Documentation}}}
  (\bibinfo {year} {2021})\BibitemShut {NoStop}%
\end{thebibliography}%

\end{document}